\documentclass[final,5p,times,twocolumn,sort&compress]{elsarticle}
\usepackage{amssymb}
\usepackage{color}
\usepackage[dvipsnames]{xcolor}
\colorlet{orange}{orange!120!}
\usepackage{graphicx}
\usepackage{dcolumn}
\usepackage{bm}
\usepackage{amstext}
\usepackage{verbatim}
\usepackage[colorlinks,citecolor=blue,linkcolor=blue,anchorcolor=blue,filecolor=blue,urlcolor=blue]{
hyperref}
\usepackage{tabularx}
\usepackage{dcolumn} 
\usepackage{ulem} 
\usepackage{longtable}
\usepackage{soul}
\newcolumntype{d}[1]{D{.}{.}{#1}} 
\definecolor{americanrose}{rgb}{1.0, 0.01, 0.24}

\journal{Physics Letters B}

\begin{document}
\begin{frontmatter}
\title{GW170817 constraints analyzed with Gogny forces and momentum-dependent interactions}

\author{O. Louren\c{c}o$^1$}
\author{M. Bhuyan$^{2,3}$}
\author{C. H. Lenzi$^1$}
\author{M. Dutra$^1$}
\author{C. Gonzalez-Boquera$^4$}
\author{M. Centelles$^4$}
\author{X. Vi\~nas$^4$}
\address{\mbox{$^1$Departamento de F\'isica, Instituto Tecnol\'ogico de Aeron\'autica, DCTA, 
12228-900, S\~ao Jos\'e dos Campos, SP, Brazil}\\
\mbox{$^2$Department of Physics, Faculty of Science, University of Malaya, Kuala Lumpur 50603, 
Malaysia}\\
\mbox{$^3$ Institute of Research Development, Duy Tan University, Da Nang 550000, Vietnam.}\\
\mbox{$^4$Departament de F\'isica Qu\`antica i Astrof\'isica and Institut de Ci\`encies del 
Cosmos (ICCUB), Facultat de F\'isica,} \mbox{Universitat de Barcelona, Mart\'i i Franqu\`es 1, 
E-08028 Barcelona, Spain}}

\begin{abstract}
A set of equations of state obtained from finite-range Gogny forces and momentum-dependent 
interactions is used to investigate the recent observation of gravitational waves from the binary 
neutron star merger GW170817 event. For this set of interactions, we have calculated the neutron 
star tidal deformabilities (related to the second Love number), the mass-radius diagram, and the 
moment of inertia~($I$). The $I$-Love relation has been verified. We also have found strong 
correlations among the tidal deformability of the canonical neutron star, its radius, and the 
derivatives of the nuclear symmetry energy at the saturation density. Most of the obtained results 
are located within the constraints of the tidal deformabilities extracted from the GW170817 
detection. 
\end{abstract}

\end{frontmatter}

\section{Introduction}
\label{intro}
The theoretical analysis of astronomical observations of very dense matter in the Universe has 
disclosed that various properties of neutron stars (NSs), such as the mass-radius relation, the 
moment of inertia, and the tidal deformability, are very sensitive to the properties of nuclear 
matter at saturation and also at supra-nuclear densities. Hence, it is important to study 
theoretically the core of NSs, which in the center can attain densities of several times $\rho_0$, 
where $\rho_0$ is the saturation density. Different nuclear models have since long been applied to 
analyze the properties of matter at these supra-nuclear densities through the equation of state 
(EoS), see e.g.~\cite{glend99,haak12,latt13,latt14,ozel16} and references therein. The recent LIGO 
and Virgo observation of GW170817~\cite{ligo17,ligo18,ligo19}, accounting for a merger of two NSs, 
has enhanced the present interest to examine the sensitivity of the EoS at large values of density 
and isospin asymmetry. Furthermore, the extracted NS tidal deformabilities of the binary system 
with 90\% and 50\% confidence limits, see Fig.~1 of Ref.~\cite{ligo18}, have provided a new 
constraint for the nuclear EoS. Many EoSs have been studied under $\beta$-equilibrium to formulate 
various connections between the tidal deformability (related to the second Love number) and other 
quantities~\cite{malik18,science,yagi13,yagi16,pank18,land18} as, for instance, the moment of 
inertia~($I$). This specific correlation is named as \mbox{$I$-Love} relation, and it seems to 
be universal, i.e., EoS-model independent.

The total mass is one of the best established observables of NSs from many observational studies. 
Among them, there are the recent accurate observations of highly massive NSs, corresponding to 
$(1.928 \pm 0.017)M_{\odot}$~\cite{demo10,fons16} and $(2.01 \pm 0.04)M_{\odot}$~\cite{anto13} 
for the \mbox{PSR J1614-2230} and \mbox{PSR J0348+0432} pulsars, respectively, with $M_\odot$ 
being the solar mass. As a result, a great effort has been addressed to derive nuclear models able 
to generate EoSs that predict such massive objects (see~\cite{oert17,vida14} and references 
therein). However, a precise mass measurement is not enough to completely constrain the underlying 
EoS. One would also need a precise measurement of the radius of the NS whose mass has been obtained. 
The uncertainties in the determination of the NS radius are still an open question for 
observational studies~\cite{sten14,nat16,de18}. The Neutron Star Interior Composition Explorer 
(NICER) mission is already set up with the aim to provide a measurement of the radius with accuracy 
of order 5\%. Moreover, the recent GW170817 observation of a NS merger~\cite{ligo17,ligo18,ligo19} 
has dramatically changed the present status of the nuclear matter models by adding new constraints. 

For several decades, a large amount of theoretical studies have been performed using different 
models to describe strongly interacting cold matter at high baryon densities in 
NSs~\cite{de18,anna18,fatt18,most18}. Very recently, many hadronic matter models have been tested 
and tried to constrain their outcomes with the observed tidal boost in the gravitational wave (GW) 
detection from the binary NS merger GW170817~\cite{ligo17,ligo18,ligo19}. Among these studies, 
nonrelativistic models~\cite{malik18,glend85,gand12,pasc18,skyrme}, effective field theories 
\cite{hebe11,kais12}, and the relativistic mean-field description of nucleons interacting via meson 
exchange~\cite{malik18,land18,fatt18}, have significantly contributed to correlate nuclear 
observables at saturation density with the cold nuclear matter at very high densities. On the other 
hand, studies based on effective quark models are focused on the existence of a possible quark 
matter phase and on the role of color superconductivity at high 
densities~\cite{oert17,buba05,beni15,fuku11,alfo98,rane16}. Other recent studies 
relating the tidal deformability to dense matter properties can also be found, for instance, in 
Refs.~\cite{tews18,perot,tsang19,apj}. Our aim in this work is to try to understand 
the behavior of NS matter at high densities using finite-range interactions of Gogny 
type~\cite{gogny1,gogny2} and of momentum-dependent interaction~(MDI) type~\cite{das03,cc2}, which 
are very successful in nuclear physics as we point out below. Our investigations will focus on the 
examination of the applicability of the presented interactions in describing the recent data on the 
GW170817 event~\cite{ligo18}, on the search for possible correlations between matter bulk parameters 
and the tidal deformability, and on the verification of the $I$-Love relation. 

The paper is organized as follows. Sec. \ref{gmdi} gives a brief description for the Gogny and 
MDI interactions applied to stellar matter. The results of our calculations are presented in Sec. 
\ref{results}. Sec. \ref{summ} includes a short summary and the concluding remarks.

\section{Gogny forces and MDI interactions for stellar matter}
\label{gmdi}

The Gogny forces were introduced by D. Gogny in order to describe the mean field and the pairing 
field within the same interaction. They consist of a density-dependent zero-range term and two 
finite-range Gaussian terms that generate the momentum dependence of the Gogny potential. Gogny 
forces are well adapted for describing the ground-state systematics as well as deformation and 
excitation properties of finite nuclei. An exhaustive compilation of nuclear properties computed 
with the D1S Gogny force~\cite{d1s} can be found in~\cite{cea}. New Gogny forces such as 
D1N~\cite{d1n} and the highly accurate D1M~\cite{d1m} have been proposed. D1N and D1M take into 
account the microscopic neutron matter EoS of Friedman and Pandharipande~\cite{Friedman81} in the 
fit of their parameters and improve on the description of isovector properties. It has been found, 
however, that when one applies the usual parametrizations of the D1 family to studies of NSs, they 
can not reach masses of $\sim 2M_\odot$. To remedy this situation, a reparametrization of the D1M 
force called D1M$^*$, which predicts maximum masses of $2M_{\odot}$ when the associated EoS is used 
to solve the Tolman-Oppenheimer-Volkoff~(TOV) equations, has been formulated very 
recently~\cite{gogny2}. The Gogny parametrizations chosen to be analyzed here are those studied in 
Ref.~\cite{gogny2}, namely, the D1M, D1M* and D2 forces. D2~\cite{d2} is another recent Gogny 
interaction where the density-dependent contact term has been replaced by a finite-range term, and 
which also delivers NS masses of $2M_\odot$. Besides, we include in the study the D1M** force 
\cite{d1mss}, which is fitted similarly to D1M* but it is required to predict a NS maximum mass of 
$1.91 M_{\odot}$, corresponding to the lowest bound of the astronomical observations.

The MDI interactions~\cite{das03} have been extensively used to study heavy-ion 
collisions~\cite{das03,li08}. Similar to the Gogny case, the MDI force may be expressed as a 
zero-range contribution plus a finite-range term, but with a Yukawa form factor instead of a 
Gaussian. A parameter $x$ of the MDI interaction can be varied to modify the uncertain density 
dependence of the symmetry energy and of the neutron matter EoS, without changing the EoS of 
symmetric nuclear matter and the symmetry energy at the saturation density. The MDI model with the 
tuned isospin dependence has also been used to describe hot asymmetric matter~\cite{li08} and 
the properties of NSs~\cite{cc2} and most recently it has been applied to study the tidal 
deformability of NSs~\cite{kras19}. In our study, we select MDI parametrizations with $-1\leqslant x 
\leqslant 0.2$. This leads to a symmetry energy slope $L_0$ at the saturation density (namely, $L_0 
= 3 \rho_0 \, \partial E_\mathrm{sym} (\rho)/ \partial \rho |_{\rho_0}$) in the range of 
$51\mbox{ MeV}\leqslant L_0 \leqslant 106\mbox{ MeV}$ for the MDI forces, whereas the range 
predicted by the Gogny parametrizations used here is $25\mbox{ MeV}\leqslant L_0 \leqslant 45\mbox{ 
MeV}$ (see Table~\ref{tabns} below). One can verify that these two boundaries for $L_0$ are 
compatible with those found in Ref.~\cite{lplb} ($L_0=58.9 \pm 16.5$~MeV) and used in 
Ref.~\cite{rmf}, in Ref.~\cite{limlatt} ($40.5\mbox{ MeV}\leqslant L_0 \leqslant 61.9\mbox{ MeV} 
$), in Ref.~\cite{oert17} \mbox{($L_0=58.7 \pm28.1$~MeV)}, and in Ref.~\cite{margueron} 
($L_0=60\pm15$~MeV). We note, however, that some of the considered MDI forces have a slope 
parameter $L_0$ above the upper boundary suggested in these references.

In stellar matter under charge neutrality and $\beta$-equilibrium, the weak process and its inverse 
reaction, namely, $n\rightarrow p + e^- + \bar{\nu}_e$ and $p+ e^-\rightarrow n + \nu_e$, take place 
simultaneously. At densities for which the electron chemical potential $\mathbf{\mu_e}$ exceeds the 
muon mass ($m_\mu$), i.e., $\mu_e > m_\mu$, the appearance of muons in the system is energetically 
favorable. By considering only these two types of leptons in the NS matter, since we assume that 
neutrinos can escape due to their extremely small cross-sections, one can write the total energy 
density and pressure of the system as $\epsilon=\epsilon_{had}+\sum_l\epsilon_l$ and $p=p_{had} + 
\sum_lp_l$, respectively, where the indexes $had$ and $l$ stand for the hadrons and leptons. In this 
work the hadronic quantities are calculated using the aforementioned Gogny and MDI interactions, 
taking into account the following conditions: $\mu_n - \mu_p = \mu_e=\mu_\mu$ and $\rho_p - \rho_e  
= \rho_\mu$, where $\rho_p=y_p\rho$, and \mbox{$\rho_l=[(\mu_l^2 - m_l^2)^{3/2}]/(3\pi^2)$} 
for $l=e, \mu$ (we use the physical values for the electron mass $m_e$ and the muon mass $m_\mu$). 
The chemical potentials and densities of protons, neutrons, electrons and muons are given, 
respectively, by $\mu_p$, $\mu_n$, $\mu_e$, $\mu_\mu$, and $\rho_p$, $\rho_n$, $\rho_e$, $\rho_\mu$, 
whereas $y_p$ denotes the proton fraction of the system. In this work we are 
dealing with NSs old enough to assume that neutrinos have left the star and therefore the 
$\beta$-equilibrium is established between the nucleons, electrons, and muons.

In order to describe a spherically symmetric NS of mass~$M$, we solve the TOV 
equations~\cite{tov39,tov39a} given (in units of $G=c=1$) by  $dp(r)/dr=-[\epsilon(r) + p(r)][m(r) 
+ 4\pi r^3p(r)]/r^2f(r)$ and $dm(r)/dr=4\pi r^2\epsilon(r)$, where $f(r)=1-2m(r)/r$. The solution 
is constrained to $p(0)=p_c$ (central pressure) and $m(0) = 0$. At the star surface, one has $p(R) = 
0$ and $m(R)\equiv M$, with $R$ defining the NS radius. To describe the EoS of the matter in the NS 
core we use the Gogny and MDI interactions. For the NS crust we consider two regions, the outer and 
the inner crust. For the former, we use the EoS proposed by Baym, Pethick and Sutherland 
(BPS)~\cite{bps} in a density region of 
$6.3\times10^{-12}\,\mbox{fm}^{-3}\leqslant\rho_{\mbox{\tiny 
outer}}\leqslant2.5\times10^{-4}\,\mbox{fm}^ {-3}$. Currently, microscopic calculations of the 
EOS of the inner crust for the different Gogny and MDI interactions are not available. Here,
following previous literature~\cite{poly0,poly1,poly2, gogny2, cc2, gogny1}, we use for the inner 
crust a polytropic EoS of the form $p(\epsilon)=A+B\epsilon^{4/3}$. The index $4/3$ assumes 
that the pressure at these densities is dominated by the relativistic degenerate electrons. For 
each interaction, we match this polytropic formula continuously to the BPS EoS at the interface 
between the outer and the inner crust and to the EoS of the homogeneous core at the core-crust 
transition pressure and energy density computed with the thermodynamical method~\cite{gogny1,cc2, 
kubis04, gonzalez19}. Thus, the core-crust matching occurs at the transition point that is predicted 
by each nuclear model. A similar procedure of description of the stellar matter EoS 
was performed in Ref.~\cite{tsang19}. In this work, the authors analyzed more than 200 Skyrme 
parametrizations in the light of GW170817, also applying a relativistic Fermi gas EoS for the inner 
crust as in our case. Another approach used to describe stellar matter is the unified treatment such 
as the one developed in Ref.~\cite{fortin} in which possible thermodynamical inconsistencies are 
avoided. In our calculations, with the present models and with some models where the unified EoS is 
available, we verified by using different crust prescriptions that the crustal EoS has no 
significant influence on the results for the tidal deformability $\Lambda$. The same conclusion has 
been brought into notice recently \cite{poly2}.

\section{Application to the GW170817 constraints}
\label{results}

In a binary NS system, tidal forces originated from the gravitational field induce tidal 
deformabilities in each companion star analogously to the tides generated on Earth due to the Moon. 
Deformations in the stars related to the quadrupole moment generate GW in which the phase evolution 
depends on the tidal deformability~\cite{tanj10,read,pozzo}. In a recent work~\cite{ligo17}, the 
LIGO/Virgo Collaboration (LVC) published an analysis on the first detection of GW coming from a 
binary NS inspiral (the energy flux out of the binary due to the GW emission causes the 
inspiral motion). The measured data allowed the LVC to determine constraints on the dimensionless 
tidal deformabilities $\Lambda_1$ and $\Lambda_2$ for each NS in the binary system, as well as on 
that one related to a canonical star of $1.4 M_\odot$ ($\Lambda_{1.4}$). Later on, in 
Refs.~\cite{ligo18,ligo19}, the constraints on the $\Lambda_1\times\Lambda_2$ region and in the 
$\Lambda_{1.4}$ value were updated. Here we test the Gogny and MDI interactions against the LVC 
constraints. In order to do that, we calculate the dimensionless tidal deformability as $\Lambda = 
2k_2/(3C^5)$, with the NS compactness given by $C=M/R$ and the second Love number written as
\begin{eqnarray}
k_2 &=&\frac{8C^5}{5}(1-2C)^2[2+2C(y_R-1)-y_R]\nonumber\\
&\times&\Big\{2C [6-3y_R+3C(5y_R-8)] \nonumber\\
&+& 4C^3[13-11y_R+C(3y_R-2) + 2C^2(1+y_R)]\nonumber\\
&+& 3(1-2C)^2[2-y_R+2C(y_R-1)]{\rm ln}(1-2C)\Big\}^{-1},\,\,
\label{k2}
\end{eqnarray}
with $y_R\equiv y(R)$, where $y(r)$ is obtained through the solution of $r(dy/dr) + y^2 + yF(r) 
+ r^2Q(r)=0$, that has to 
be solved as part of a coupled system containing the TOV equations. The quantities $F(r)$ and 
$Q(r)$ are defined as
\begin{eqnarray}
F(r) &=& \frac{1 - 4\pi r^2[\epsilon(r) - p(r)]}{f(r)} , 
\\
Q(r)&=&\frac{4\pi}{f(r)}\left[5\epsilon(r) + 9p(r) + 
\frac{\epsilon(r)+p(r)}{v_s^2(r)}- \frac{6}{4\pi r^2}\right]
\nonumber\\ 
&-& 4\left[ \frac{m(r)+4\pi r^3 p(r)}{r^2f(r)} \right]^2,
\label{qr}
\end{eqnarray}
where $v_s^2(r)=\partial p(r)/\partial\varepsilon(r)$ is the squared sound velocity.
See for instance Refs.~\cite{tanj10,new,hind08,damour,tayl09} for the derivations.

\begin{figure}[!t]
\centering
\includegraphics[scale=0.3]{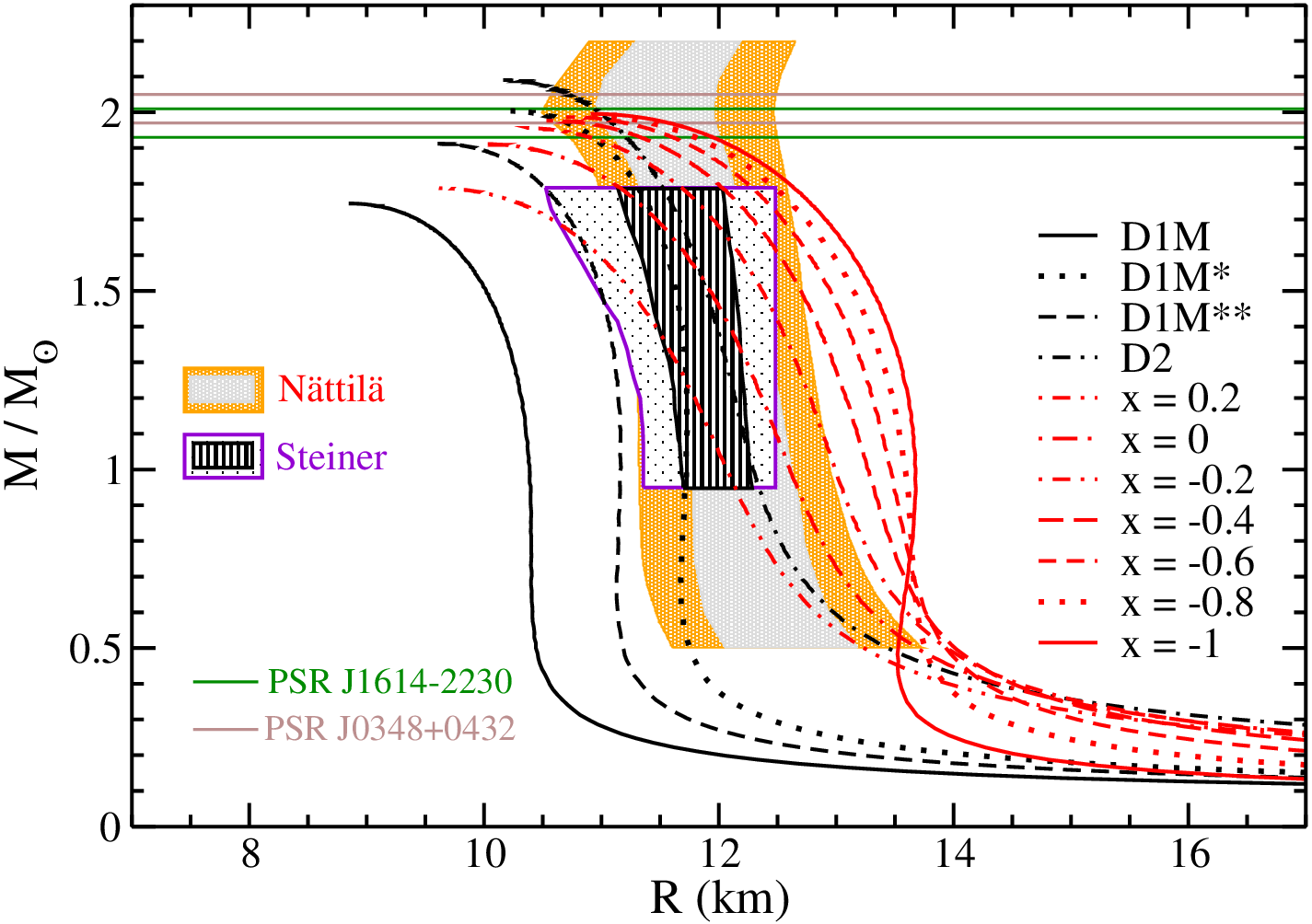}
\caption{Neutron star mass-radius diagrams for Gogny and MDI parametrizations. Horizontal bands
indicate the masses of \mbox{PSR~J1614-2230}~\cite{demo10} and \mbox{PSR~J0348+0432}~\cite{anto13} 
pulsars. Outer orange and inner grey bands: constraints extracted from Ref.~\cite{nat16}. 
Outer white and inner black bands: constraints extracted from Ref.~\cite{stein10}.}
\label{mr}
\end{figure}
\begin{table}[!tbh]
\centering
\caption{Stellar matter properties obtained from the Gogny and MDI models used here: maximum 
neutron star mass~($M_{\rm max}$) with its corresponding radius~($R_{\rm max}$) and central energy 
density~($\epsilon_c$), along with the radius~($R_{1.4}$) of a canonical neutron star of mass of 
$M=1.4M_\odot$, and its dimensionless tidal deformability~($\Lambda_{1.4}$). The value of the slope 
parameter $L_0$ of the symmetry energy for the considered models is shown in the last column.}
\resizebox{\columnwidth}{!}{%
\begin{tabular}{lccccccccc}
\hline
Model & $M_{\rm max}$ & $R_{\rm max}$ & $\varepsilon_c$ & $R_{\rm 1.4}$ & 
$\Lambda_{\rm 1.4}$ & $L_0$ \\
          &  ($M_\odot$)  & (km)          & (fm$^{-4}$)     & (km)          & & (MeV)\\
\hline
D1M            & 1.74 &  8.85 & 10.40 & 10.15 & 122 & 24.83 \\
D1M*           & 2.00 & 10.20 &  7.74 & 11.69 & 316 & 43.18 \\
D1M**          & 1.91 &  9.60 &  8.78 & 11.07 & 221 & 33.91 \\
D2             & 2.09 & 10.16 &  7.80 & 11.98 & 300 & 44.83 \\
MDI ($x=0.2$)  & 1.79 &  9.58 &  9.33 & 11.46 & 217 & 51.05 \\
MDI ($x=0.0$)  & 1.91 & 10.00 &  8.49 & 12.10 & 314 & 60.17 \\
MDI ($x=-0.2$) & 1.96 & 10.32 &  7.97 & 12.52 & 402 & 69.28  \\
MDI ($x=-0.4$) & 1.98 & 10.54 &  7.65 & 12.84 & 487 & 78.40  \\
MDI ($x=-0.6$) & 1.99 & 10.70 &  7.44 & 13.09 & 573 & 87.51  \\
MDI ($x=-0.8$) & 1.99 & 10.82 &  7.27 & 13.29 & 661 & 96.63  \\
MDI ($x=-1.0$) & 2.00 & 10.91 &  7.15 & 13.45 & 754 &105.75 \\
\hline
\end{tabular}
}
\label{tabns}
\end{table}
We show in Fig.~\ref{mr} the NS mass-radius diagrams for the Gogny and MDI parametrizations used in 
this work. Some quantities related to the maximum NS mass and a canonical NS are shown in 
Table~\ref{tabns}. Concerning the Gogny model, the parametrization D1M*~\cite{gogny2} is compatible 
with the bands coming from constraints derived from data obtained, with some underlying 
assumptions for the used EoS, by Steiner~\cite{stein10} and N\"attil\"a~\cite{nat16}. It also 
predicts NSs with $M\sim 2M_\odot$, in agreement with the well-known data from the \mbox{PSR 
J1614-2230}~\cite{demo10} and \mbox{PSR J0348+0432}~\cite{anto13} pulsars. These features for 
D1M$^*$ also hold for the D2 force~\cite{d2}. Concerning the MDI models, we see that the model with 
$x=0.2$ does not reach $1.9 M_\odot$, whereas $M_{\rm max}$ of the other models is above $1.9 
M_\odot$. A very recent value for the observed largest NS mass of $2.14^{+0.20}_{-0.18} M_\odot$
at 95.4\% credible level and of $2.14^{+0.10}_{-0.09} M_\odot$ at 68.3\% credible level is proposed
in Ref.~\cite{cromartie} from the \mbox{MSP J0740+6620} pulsar. D1M*, D2 and the MDI models with $x 
\leq -0.2$ would be compatible with the lower limit of this mass at the 95.4\% level and only D2 in 
the case of the 68.3\% level. 

Still, regarding the values shown in Table~\ref{tabns}, we remark that it seems there is at least a 
trend of a certain ``threshold value'' of $L_0\sim 40$~MeV, the lower limit of the range proposed 
in Ref.~\cite{limlatt}, below which there are only configurations predicting $M_{\mbox{\tiny max}}< 
2M_\odot$ for the maximum NS mass. For instance, this is true for the D1M and D1M** parametrizations 
and for the $42$ parametrizations of Skyrme and relativistic models analyzed in 
Ref.~\cite{constanca}. A more systematic study in that direction is needed in order to confirm this 
result for a larger set of models. For now on, we restrict our study only to 
parametrizations presenting $M_{\rm max}\sim 2M_\odot$, i.e., we exclude the EoSs of D1M and 
MDI($x=0.2$), which predict maximum masses clearly below the observations of 
Refs.~\cite{demo10,fons16,anto13}.

\begin{figure}[!tbh]
\centering
\includegraphics[scale=0.3]{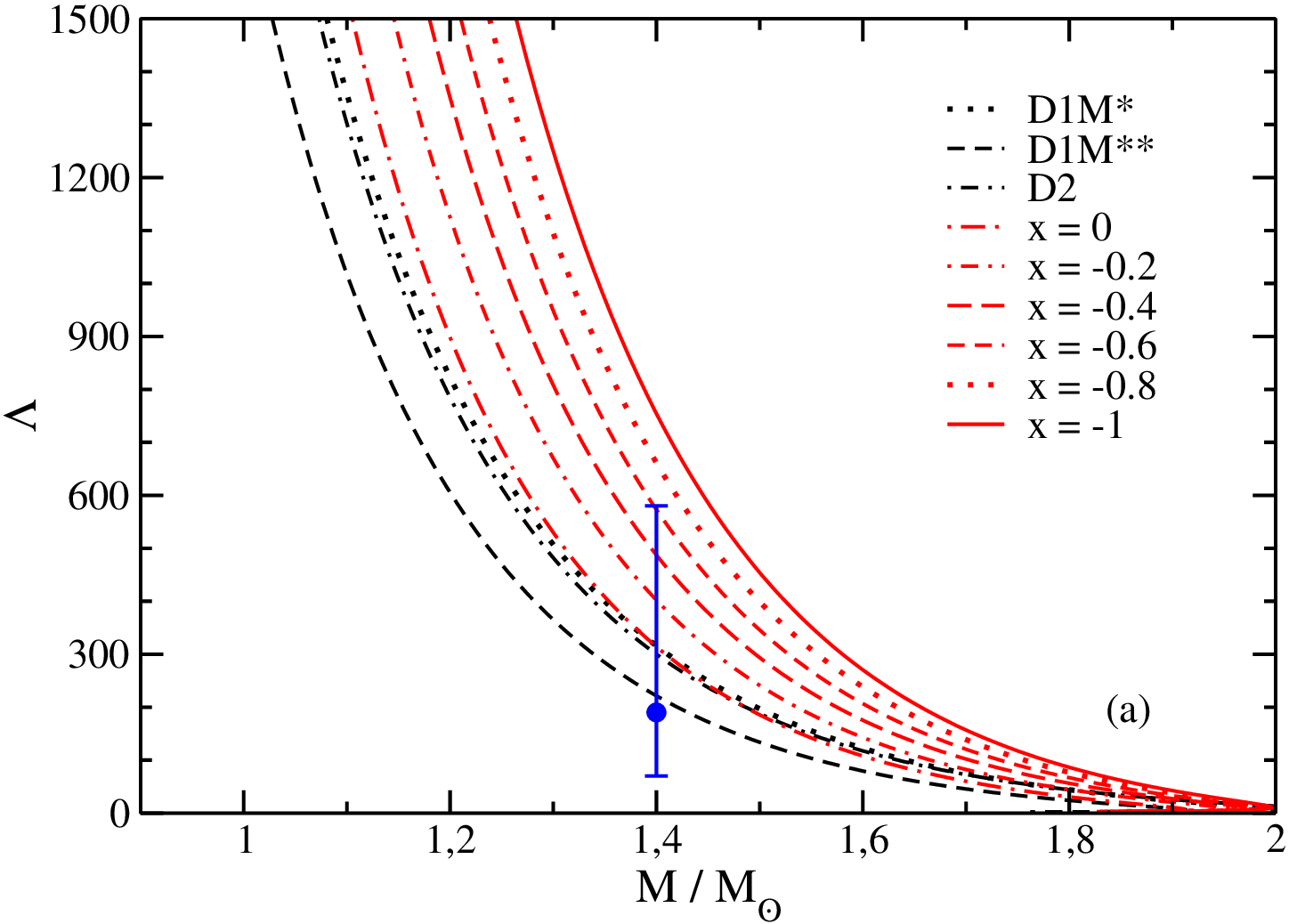}
\includegraphics[scale=0.3]{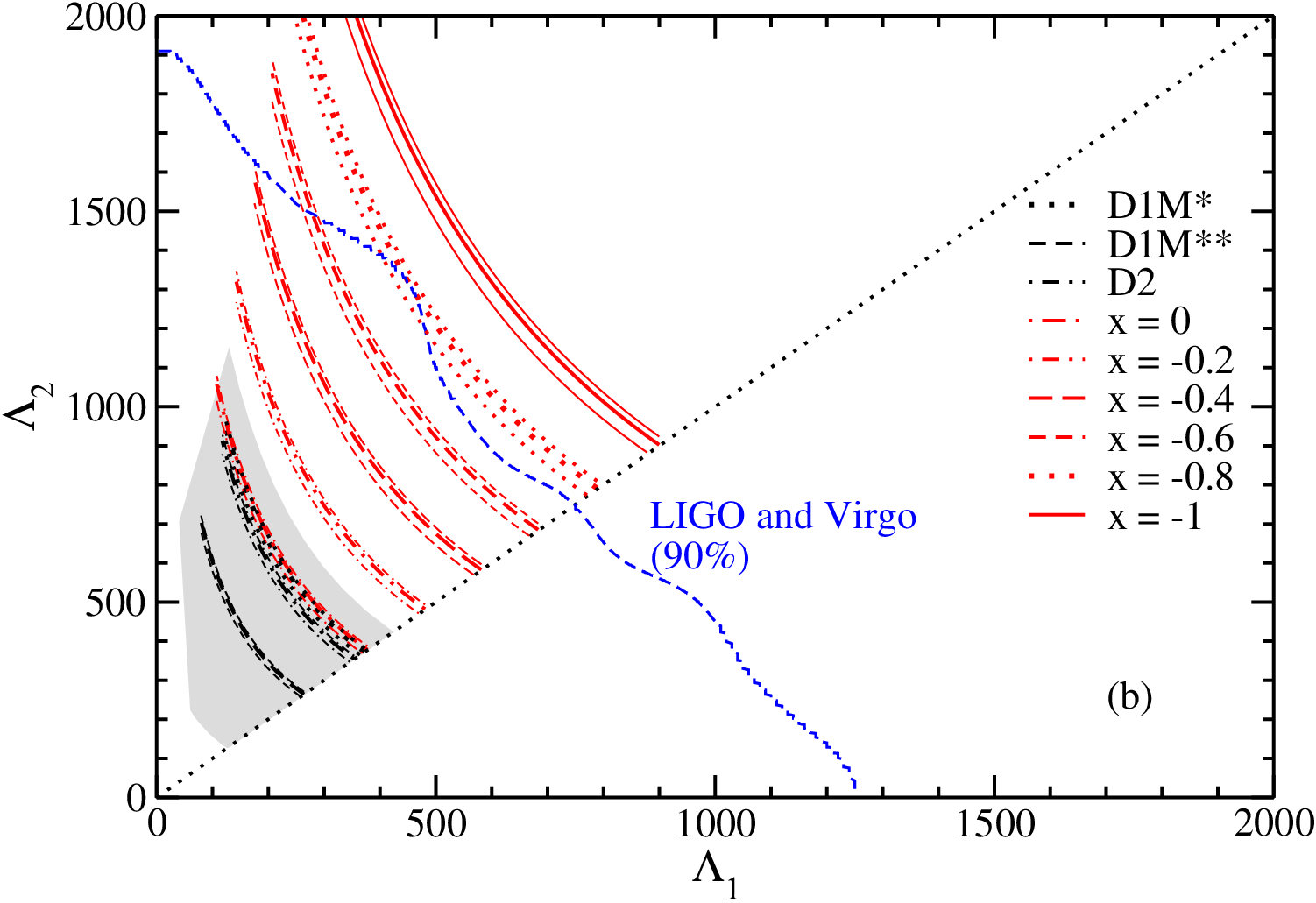}
\caption{(a) Dimensionless tidal deformability $\Lambda$ as a function of the neutron star 
mass, and (b) dimensionless tidal parameters $\Lambda_1$ and $\Lambda_2$ related 
to the high and low mass components inferred from the binary system of the GW170817 event, along 
with the contour line of $90\%$ credible level (dashed blue curve). The internal and 
external curves of each parametrization are related to the error in the chirp mass, $\mathcal{M}= 
1.188^{+0.004}_{-0.002}M_\odot$~\cite{ligo17}. Both panels are constructed 
from the Gogny and MDI parametrizations used in this work. The blue solid circle with error bars 
(panel a) is related to the constraint extracted from the GW170817 event~\cite{ligo18}. Grey band: 
predictions from Skyrme parametrizations consistent with constraints from symmetric and asymmetric 
nuclear matter~\cite{skyrme,dutra12}.} 
\label{lambda}
\end{figure}

In Fig.~\ref{lambda}{\color{blue}a} we plot the dimensionless tidal deformability~$\Lambda$ as 
a function of the NS mass. We see that all parametrizations, except the MDI ones with $x=-0.8$ and 
$x=-1$, reproduce the data from LIGO/Virgo~\cite{ligo18} of $\Lambda_{1.4}=190^{+390}_{-120}$ for 
the canonical star. Furthermore, by using the results presented in Fig.~\ref{lambda}{\color{blue}a} 
and in Table~\ref{tabns}, it is possible to extract the following fitting expression for 
$\Lambda_{1.4}$ as a function of the canonical star radius, $\Lambda_{1.4}=aR_{1.4}^b$, with 
$a=6.97\times10^{-5}$ and $b=6.19$. Although $\Lambda$ is defined as proportional to 
$k_2R^5$, $k_2$ is obtained through the solution of the differential equation for $y$ as a function 
of $R$, see Eq.~(\ref{k2}). This is the origin of the power not equal to~5 in the 
$\Lambda_{1.4}\times R_{1.4}$ relationship. Different $a$ and $b$ values are found for different 
relativistic and nonrelativistic models, as one can verify for instance in 
Refs.~\cite{malik18,anna18,fatt18}. Such a relation has also been derived in 
Ref.~\cite{tews-margueron} from a very general approach based on a parameterization of the speed of 
sound. The relation $\Lambda_{1.4}\times R_{1.4}$ is important, in particular, since there is a 
range for $\Lambda_{1.4}$ provided by the LVC, namely, $\Lambda_{1.4}=190^{+390}_{-120}$. If we 
combine this range with our fitting expression $\Lambda_{1.4}=aR_{1.4}^b$, it is possible to extract 
a respective range of $R_{1.4}$ by inverting the expression as $R_{1.4}=(\Lambda_{1.4}/a)^{1/b}$. By 
applying the boundary values of $\Lambda_{1.4}=70$ and $\Lambda_{1.4}=580$, one can extract a range 
of $9.3\mbox{ km} \le R_{1.4} \le 13.1\mbox{ km}$ for the radius of a canonical NS. This range is 
very similar to the one predicted in Ref.~\cite{tews18}, namely, $9.0\mbox{ km} \le R_{1.4} \le 
13.6\mbox{ km}$, where the authors confront the LVC data with their theoretical analysis.

In Fig.~\ref{lambda}{\color{blue}b} we show the tidal deformabilities $\Lambda_1$ and $\Lambda_2$ 
of a binary NS system calculated with the Gogny and MDI parametrizations along with the contour 
line of $90\%$ credible level (dashed blue curve) related to the GW170817 NS merger event 
\cite{ligo18}. To generate the curves, we vary the mass of one of the stars in the binary system, 
$m_1$, in the range $1.36 \leq m_1/M_\odot \leq 1.60$~\cite{ligo18,ligo19}. The mass of the 
companion star, $m_2$, is related to $m_1$ through the so-called chirp mass $\mathcal{M}$, which is 
given by ${\mathcal M} = (m_1 m_2)^{3/5}/(m_1+m_2)^{1/5}$. The GW170817 detection provides a chirp 
mass of $\mathcal{M}= 1.188^{+0.004}_{-0.002}M_\odot$~\cite{ligo17}. Hence, $m_2$ varies in the 
ranges $1.17 \leq m_2/M_\odot \leq 1.36$~\cite{ligo17,ligo18}. From Fig.~\ref{lambda}{\color{blue}b} 
we find a good agreement between the results obtained with the Gogny and MDI parametrizations used 
in this work and the LIGO/Virgo data. We also compare the present results with the values provided 
by the widely-known Skyrme model~\cite{skyrme}. In that work, the authors calculated the tidal 
deformabilities for a set of Skyrme forces that are consistent with the constraints in symmetric and 
asymmetric nuclear matter discussed in Ref.~\cite{dutra12}. We see that all the Gogny forces 
analyzed here lie inside the region defined by the Skyrme models, which is represented  by the grey 
band in Fig.~\ref{lambda}{\color{blue}b}. A similar kind of agreement is found for MDI with $x=0$ 
and $x=0.2$.

\begin{figure}[!t]
\centering
\includegraphics[scale=0.35]{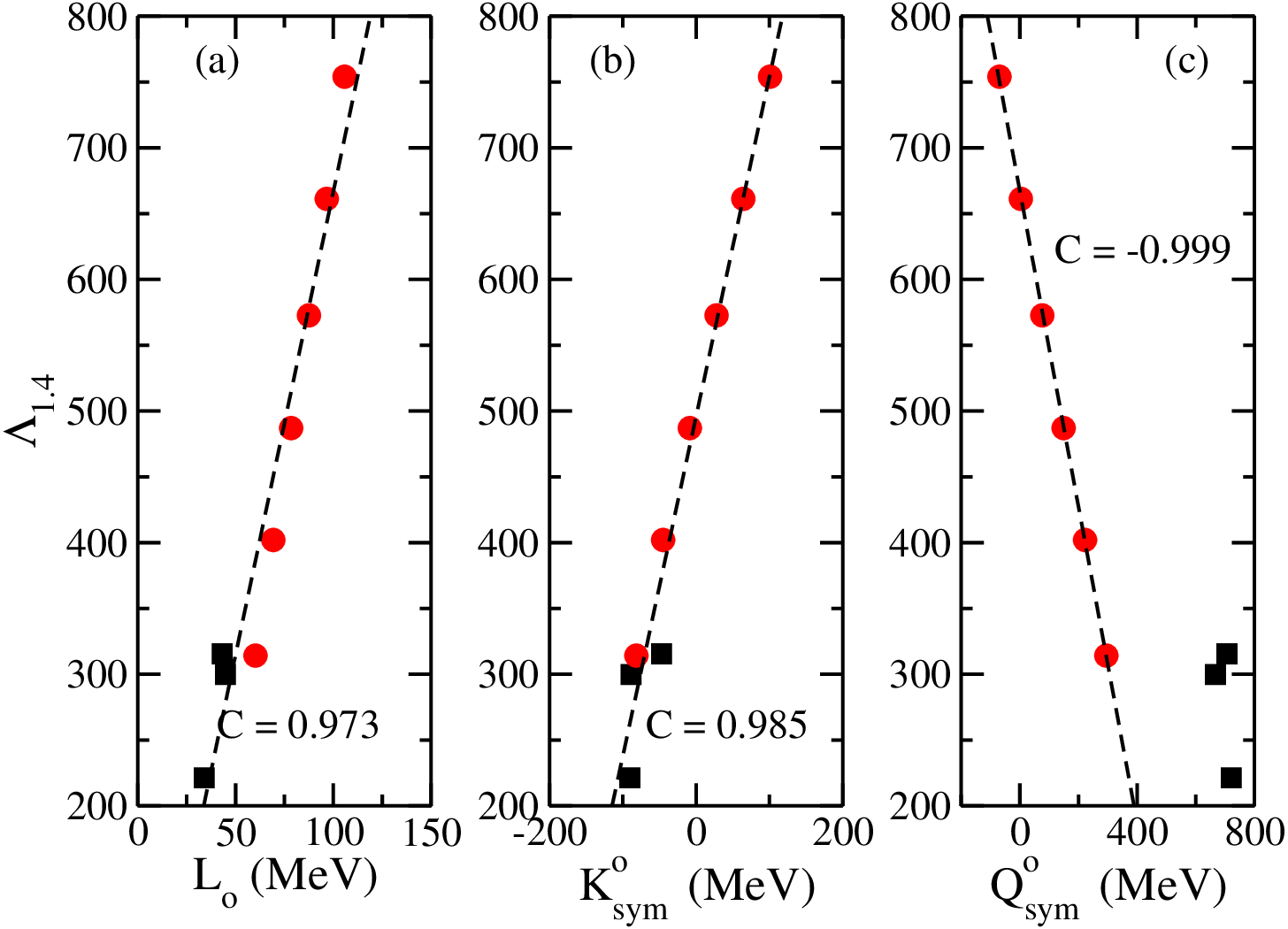}
\caption{Results from Gogny (black squares) and MDI (red circles) models for $\Lambda_{1.4}$ 
as a function of (a) symmetry energy slope, (b) symmetry energy curvature and (c) symmetry 
energy skewness, all of them at the saturation density. Dashed lines: fitting curves.}
\label{corr}
\end{figure}

Another investigation performed in this work is the search for possible correlations between the 
tidal deformability $\Lambda_{1.4}$ and the bulk parameters of nuclear matter. Since $\Lambda_{1.4}$ 
is constrained by the LIGO/Virgo analysis of the GW170817 NS merger event, correlations between this 
quantity and a generic nuclear quantity $\mathcal{A}$ may help to establish boundaries on 
$\mathcal{A}$, and, consequently, to better constrain the microphysics related to the EoS of the 
hadronic model. This is the case for the nuclear matter bulk 
parameters~\cite{bianca,pieka-centelles,khan-margueron}. In Figs.~\ref{corr}{\color{blue}a} 
and~\ref{corr}{\color{blue}b}, we display $\Lambda_{1.4}$ as a function of $L_0$ and 
$K^0_{\mbox{\tiny sym}}$, namely, the symmetry energy slope and energy curvature (both at the 
saturation density) with the respective fitting curves applied to the considered Gogny and MDI 
parametrizations. In Fig.~\ref{corr}{\color{blue}c} we show how $\Lambda_{1.4}$ depends on the 
symmetry energy skewness at the saturation density, $Q^0_{\mbox{\tiny sym}}$. Here, the fitting 
curve is obtained only for the MDI parametrizations, since we have not enough points related to the 
Gogny model.
From the figure, we verify a strong correlation of $\Lambda_{1.4}$ with $L_0$ and 
$K^0_{\mbox{\tiny sym}}$, as the fitted curves show a high correlation coefficient. The increasing 
of $\Lambda_{1.4}$ with $L_0$ was also verified in Ref.~\cite{perot}, in which the authors used a 
set of unified equations of state based on the Skyrme model. We find in 
Fig.~\ref{corr}{\color{blue}c} a high correlation of $\Lambda_{1.4}$ with $Q^0_{\mbox{\tiny sym}}$ 
within the MDI model. However, although there are only few Gogny points, the results suggest that 
the trend for $\Lambda_{1.4}$ with $Q^0_{\mbox{\tiny sym}}$ is different for Gogny and MDI models. 
Analyzing in addition the predictions of several Skyrme parametrizations (not shown) we have also 
observed that the correlation of $\Lambda_{1.4}$ with $Q^0_{\mbox{\tiny sym}}$ is weak among 
different models. One can assign this larger deviation to the higher-order derivative of the 
symmetry energy, i.e., small variations on the symmetry energy are enough to cause larger variations 
on $Q^0_{\mbox{\tiny sym}}$ (third derivative of the symmetry energy).

One can use the given correlations and the observational range for $\Lambda_{1.4}$ predicted in 
Ref.~\cite{ligo18} to determine compatible ranges for $L_0$ and $K^0_{\mbox{\tiny sym}}$. By 
applying this procedure we find that the dimensionless tidal deformability of the canonical neutron 
star obtained by the LIGO/Virgo collaboration, namely, $\Lambda_{1.4}=190^{+390}_{-120}$, can be 
satisfactorily described by parametrizations presenting $15\,\mbox{MeV}\,\leq\,L_0\,\leq 
88\,\mbox{MeV}$ and $-165\,\mbox{MeV}\,\leq\,K^0_{\mbox{\tiny sym}}\,\leq 33\,\mbox{MeV}$. As we 
have discussed before, interactions with slope parameter $L_0 \lesssim 40$ MeV seem to be unable to 
provide NSs of $M_\mathrm{max}$ of $2 M_\odot$. This constraint, applied to 
Fig.~\ref{corr}{\color{blue}a}, points to a lower bound of $\Lambda_{1.4} \gtrsim 244$,  which in 
turn implies a lower bound of $-98$~MeV for $K^0_{\mbox{\tiny sym}}$.

We also remark that such constraints on the nuclear matter bulk parameters obtained 
from their relation with $\Lambda_{1.4}$ are an attempt to find a general trend for the values of 
these quantities. In that sense, they are model dependent and a more systematic study including 
other kind of hadronic models is needed in order to confirm or improve these numbers. 
Concerning the range for $K^0_{\mbox{\tiny sym}}$, for instance, it is worth to mention that it is 
similar to the one suggested in Ref.~\cite{margueron}, in which the authors proposed a metamodeling 
approach to treat the EoS for dense nuclear matter, namely, $K^0_{\mbox{\tiny sym}}=-100\pm100$~MeV. 
Regarding the range of $L_0$, on the other hand, there are more stringent bounds found from 
combined analyses of astrophysical and nuclear data, see Ref.~\cite{limlatt} for instance.

All of the Gogny and MDI parametrizations analyzed in this work show similar bulk properties in 
symmetric nuclear matter, with saturation density $\rho_0 \!\sim\! 0.16$ fm$^{-3}$, energy per 
particle $E_0 \!\sim\! -16$ MeV, effective mass $m^*_0/m \!\sim\! 0.7$, incompressibility $K_0$ 
between $209$ and $225$~MeV, and skewness $Q_0$ between $-460$ and $-427$~MeV. In the isovector 
sector, we find a similar situation for the symmetry energy $E_\mathrm{sym}^0$ at saturation, which 
shows a limited variation ($E_\mathrm{sym}^0 \!=\! 28.5\,-\,31$ MeV) in the considered 
parametrizations. Therefore, this fact can be the origin of the very small interference of these 
bulk parameters in possible correlations arising from $\Lambda_{1.4}$ as a function of $L_0$, 
$K^0_{\mbox{\tiny sym}}$ and $Q^0_{\mbox{\tiny sym}}$. 

\begin{figure}[!t]
\centering
\includegraphics[scale=0.35]{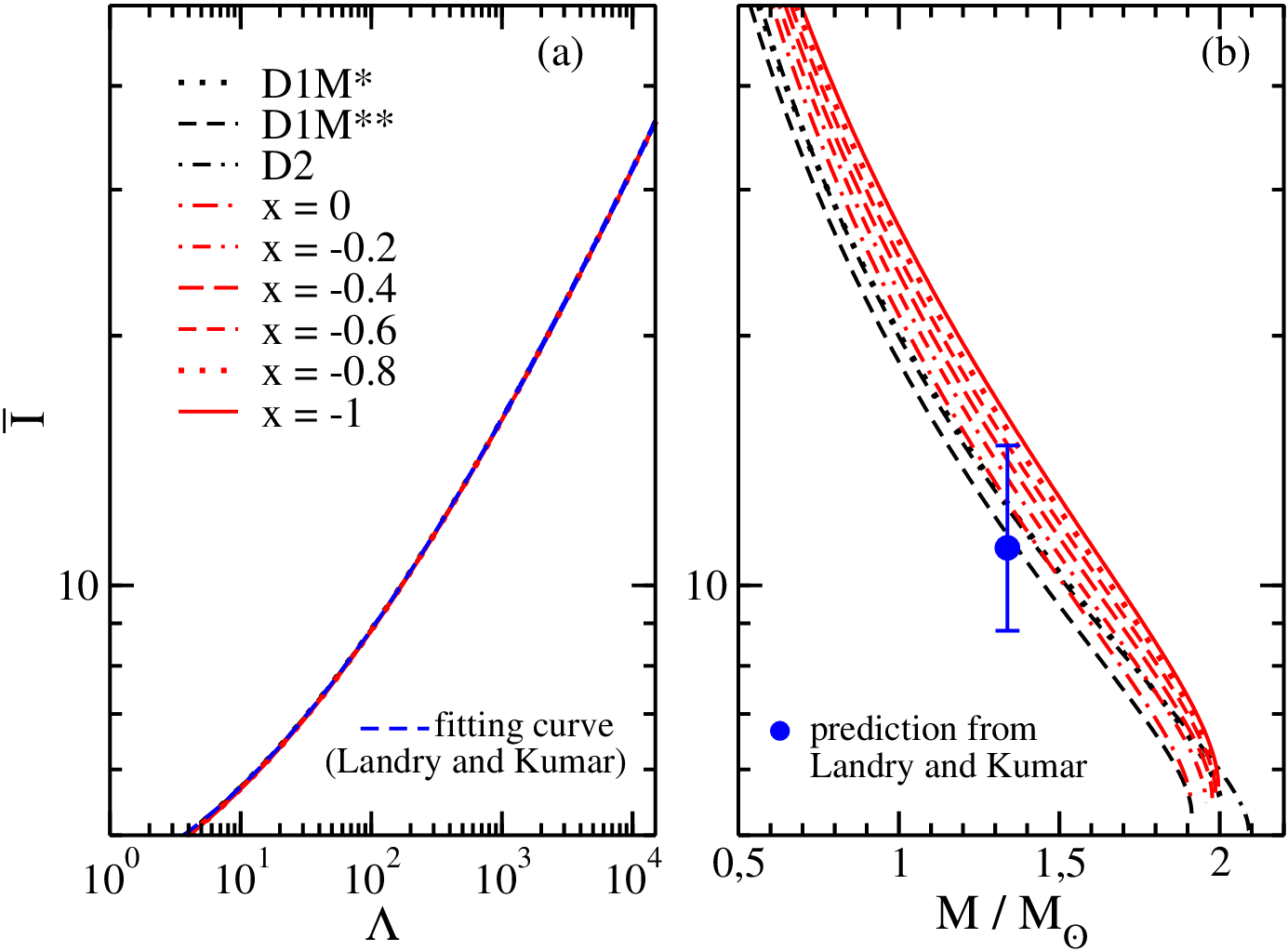}
\caption{Results from Gogny and MDI models for the dimensionless moment of inertia $\bar{I}$ as a 
function of (a)~dimensionless tidal deformability, and (b)~neutron star mass. Dashed blue curve: 
fitting curve obtained in Ref.~\cite{land18}. Blue circle with error bars: predictions from 
Ref.~\cite{land18} for $\bar{I}_\star$, namely, the dimensionless moment of inertia of the 
\mbox{PSR J0737-3039} primary component pulsar.}
\label{inertia}
\end{figure}

Finally, we show in Fig.~\ref{inertia} the dimensionless moment of inertia, $\bar{I}\equiv I/M^3$, 
calculated from the Gogny and MDI models. We obtain this quantity by solving the Hartle's slow 
rotation equation given in Refs.~\cite{land18,hartle,yagi13}, namely, a differential equation for 
one of the metric decomposition functions~\cite{yagi13}, $\omega(r)$, coupled to the TOV equations. 
The moment of inertia is given in terms of $\omega_R\equiv \omega(R)$ as $I=R^3(1-\omega_R)/2$, 
where $\omega_R$ is the frame-dragging function $\omega(r)$~\cite{land18} evaluated at 
the surface of the star in which $r=R$.
In Fig.~\ref{inertia}{\color{blue}a}, one can verify that all used Gogny and MDI interactions are 
indistinguishable in the plot of $\bar{I}$ as a function of $\Lambda$. Such a feature is claimed to 
be universal and is known as the \mbox{$I$-Love} relation. In Refs.~\cite{science,yagi13}, using a 
set of~$10$ different EoSs, the authors showed that such a relation is independent of the neutron 
star (or quark star) structure for slowly-rotating stars. In Ref.~\cite{land18}, the same result was 
obtained for a set of~$53$ Skyrme and relativistic mean-field parametrizations. Here, we show that 
the same feature remains for the Gogny and MDI finite-range nuclear interactions, thereby confirming 
the universality of the $I$-Love relation also with this class of models. In 
Fig.~\ref{inertia}{\color{blue}a} we display by a dashed line the fitting curve determined in 
Ref.~\cite{land18}. We see that the Gogny and MDI models are compatible with this fitting.

In Ref.~\cite{land18}, the authors also determined a range for $\bar{I}$ related to the \mbox{PSR 
J0737-3039} primary component pulsar (a pulsar of mass $M_\star=1.338M_\odot$), 
namely, $\bar{I}_\star\equiv\bar{I}(M_\star)=11.10^{+3.68}_{-2.28}$. Measurements regarding this 
slowly rotating pulsar are expected to determine its moment of inertia in the next few years with 
hitherto unachieved accuracy ($\sim$10\%). Such a range of Ref.~\cite{land18} was calculated, using 
the candidate Skyrme and relativistic mean-field EoSs, in two steps. First, the authors of 
\cite{land18} verified a relation between $\Lambda_\star$ and $\Lambda_{1.4}$, called as the 
\mbox{binary-Love} relation. Here, $\Lambda_\star$ is the dimensionless tidal deformability 
related to the star of mass $M_\star$. Then, they used a suitable fitting for $\Lambda_\star$ as a 
function of $\Lambda_{1.4}$ along with the \mbox{$I$-Love} relation fitting curve presented in 
Fig.~\ref{inertia}{\color{blue}a} to establish $\bar{I}_\star$ as a function of~$\Lambda_\star$. 
Finally, the observational range $\Lambda_{1.4}=190^{+390}_{-120}$ obtained by the LIGO/Virgo 
collaboration was used to determine a range for~$\Lambda_\star$, and consequently, the limits of 
$\bar{I}_\star=11.10^{+3.68}_{-2.28}$. As one can see in Fig.~\ref{inertia}{\color{blue}b}, the 
Gogny parametrizations used present $\bar{I}_\star$ inside the predicted range. For the MDI 
interactions, those in which $x>-0.6$ are also in agreement with the limits. The fact that the Gogny 
and MDI forces that fulfill the LVC observational bounds $\Lambda_{1.4}=190^{+390}_{-120}$ also 
agree with the range of Ref.~\cite{land18} for $\bar{I}$, corroborates with another type of hadronic 
models, the limits for the moment of inertia of pulsar A of \mbox{PSR J0737-3039} established by 
Landry and Kumar~\cite{land18}.

\section{Summary and concluding remarks}
\label{summ}

We have studied the predictions of the EoS of stellar matter from finite-range effective 
interactions of the Gogny and MDI types for describing NS properties related with the information 
extracted from the GW170817 detection of gravitational waves by the LIGO/Virgo collaboration. The 
tidal deformability $\Lambda_{1.4}$ predicted by the Gogny and MDI interactions for a $1.4M_\odot$ 
NS agrees with the corresponding data supplied by LIGO/Virgo, except for the MDI interactions with 
$x=-0.8$ and $x=-1$, which have the stiffest symmetry energies of the considered 
models. From the dependence of $\Lambda_{1.4}$ with the NS radius 
$R_{1.4}$ obtained with the Gogny and MDI interactions, the LVC observational value 
$\Lambda_{1.4}=190^{+390}_{-120}$ allows one to constrain the radius of a canonical NS in the range 
of $9.3\mbox{ km} \le R_{1.4} \le 13.1$~km. We have also calculated the tidal 
deformabilities of the two NSs in the binary system assuming the chirp mass that was measured in the 
GW170817 event. The results lie within the 90\% confidence region extracted from the LIGO/Virgo data 
for most of the analyzed parametrizations.

We have seen that the isoscalar properties and the value of the symmetry energy at saturation of 
the underlying Gogny and MDI interactions used to build the stellar EoS have little impact on the 
tidal deformability $\Lambda_{1.4}$ computed with these models, while $\Lambda_{1.4}$ shows a clear 
dependence, roughly linear, with the isovector properties of the interaction, such as the slope 
($L_0$), curvature ($K^0_{\mbox{\tiny sym}}$) and skewness ($Q^0_{\mbox{\tiny sym}}$) of the 
symmetry energy. We find that the LIGO/Virgo determination of $\Lambda_{1.4}$ can be reproduced by 
the effective interactions characterized by $15\,\mbox{MeV}\,\leq\,L_0\,\leq 
88\,\mbox{MeV}$ and $-165\,\mbox{MeV}\,\leq\,K^0_{\mbox{\tiny sym}}\,\leq 
33\,\mbox{MeV}$. In particular, the~$L_0$ range is in harmony with other estimates of the slope of 
the symmetry energy derived from nuclear observables such as the neutron skin thickness, giant 
resonances, heavy-ion reactions, or the systematics of nuclear masses (see, for instance, 
\cite{vinas14} and references therein). Finally, we have analyzed the moment of inertia of a NS as a 
function of the tidal deformability. We have shown that the so-called $I$-Love relation 
\cite{science,yagi13}, which is expected to be universal, is, indeed, also satisfied by the Gogny 
and MDI interactions. In addition, we obtain that the prediction of Landry and Kumar \cite{land18} 
of a dimensionless moment of inertia $\bar{I}_\star=11.10^{+3.68}_{-2.28}$ for \mbox{PSR J0737-3039} 
is supported by the Gogny and MDI models that, at the same time, are consistent with the LIGO/Virgo 
determination of $\Lambda_{1.4}$.

\section*{Acknowledgments}
This work is a part of the project INCT-FNA Proc. No. 464898/2014-5, partially supported by 
Conselho Nacional de Desenvolvimento Cient\'ifico e Tecnol\'ogico (CNPq) under grants 310242/2017-7 
and 406958/2018-1 (O.L.), and 433369/2018-3 (M.D.), and by Funda\c{c}\~ao de Amparo \`a Pesquisa do 
Estado de S\~ao Paulo (FAPESP) under the thematic projects 2013/26258-4 (O.L.) and 2017/05660-0 
(O.L., M.D.). C.G, M.C. and X.V. acknowledge support from Grant FIS2017-87534-P from MINECO and 
FEDER, and Project MDM-2014-0369 of ICCUB (Unidad de Excelencia Mar\'{\i}a de Maeztu) from MINECO.
C.G. also acknowledges Grant BES-2015-074210 from MINECO.



\end{document}